This work has been submitted to IEEE Power Engineering Letters for possible publication. Copyright may be transferred without notice, after which this version may no longer be accessible.

# Does Synchrophasor Attenuation Preclude Detection of High-Frequency Oscillations?

Chetan Mishra, *Member, IEEE*, Luigi Vanfretti, *Senior Member, IEEE*,
Jaime Delaree Jr., Member, IEEE, and Kevin D. Jones, *Member, IEEE*

***Abstract*— Recent work suggests that standard synchrophasors are inadequate for capturing high-frequency data center oscillations, primarily due to filtering and windowing attenuation during phasor estimation. We prove that widely used periodogram and RMS energy detectors remain largely robust to it and confirm this by simulation.**

## I. Introduction

Modern grids are facing high-frequency oscillations driven by power electronics [1], [2], and synchrophasors can play a critical role in tracking such dynamics. However, standard discrete Fourier transform (DFT)-based phasor estimation algorithms inherently filter the measurements, shifting phases and attenuating magnitudes, which the current literature suggests limits their ability to faithfully capture high-frequency oscillations [3], [4], raising the question of whether such oscillations can still be reliably detected. In this work, we derive the impact of DFT attenuation on the statistical performance of widely used oscillation detectors [5], [6], and prove that they are largely robust to it.

## II. DFT Attenuation Impact on Synchrophasor Magnitude Periodogram

Let the actual power system signal (voltage or current) be characterized by its 60 Hz amplitude $A(t) = T_A(t) + \epsilon_A(t)$ and phase $\phi(t) = T_\phi(t) + \epsilon_\phi(t)$, where $T\ (t)$ represents the DC component + slow trend $< 0.1\,Hz$ and $\epsilon\ (t)$ comprises stationary ambient noise and periodic components [7]. Its DFT-based synchrophasor estimate $A_p(t)\angle\phi_p(t)$ at a sampling interval of $\Delta t_w$, using a window $g(t)$ and assuming small ambient angle dynamics $\epsilon_\phi(t), \epsilon_A(t)$ can be approximated as,

$$A_p[m]e^{j\phi_p[m]} \approx e^{jT_\phi[m]}\sum_k (T_A[k]+\epsilon_A[k])e^{j\epsilon_\phi[k]}g[m-k]\Delta t_w \quad (1)$$

where $m$ denotes the waveform sampling index, i.e., $x[m] = x(m\Delta t_w)$. Finally, by taking the magnitude and only retaining the first-order terms in ambient dynamics,

$$A_p[m] \approx \left|\sum_k (T_A[k]+\epsilon_A[k])e^{j\epsilon_\phi[k]}g[m-k]\Delta t_w\right| \approx \underbrace{T_A[m]}_{T_{A_p}[m]} + \underbrace{\sum_k \epsilon_A[k]g[m-k]\Delta t_w}_{\epsilon_{A_p}[m]} \quad (2)$$

The above estimate is filtered and downsampled to a slower rate $f_s = \frac{1}{\Delta t}$ (30 or 60 Hz), with downsampling factor $M = \frac{\Delta t}{\Delta t_w}$. Oscillation detection is therefore performed using $A_p(n\Delta t) = A_p[nM]$. Oscillation detectors exploit the property that a periodic component at a specific frequency concentrates power into the corresponding frequency bin. Consequently, widely adopted oscillation detection algorithms [5], [6], [7] rely on the periodogram [8]. Given a finite detrended phasor measurement unit (PMU) magnitude data window of length $N$, the periodogram ordinate $I_N^{A_p}(f_i)$ [8] evaluated at the $i^{th}$ DFT bin, corresponding to the frequency $f_i = \frac{i}{N}f_s$ , is defined as:

$$I_N^{A_p}(f_i) = \left|\frac{\sum_{n=0}^{N-1}\epsilon_{A_p}[nM]e^{-j2\pi i\frac{n}{N}}}{\sqrt{N}}\right|^2 \quad (3)$$

Substituting the expression from (2) and swapping the summation order yields,

$$= \left|\frac{1}{\sqrt{N}}\sum_k \epsilon_A[k]\left(\sum_{n=0}^{N-1} g[nM-k]e^{-j2\pi i\frac{n}{N}}\Delta t_w\right)\right|^2 \quad (4)$$

Now, for large $N$, the inner summation $\sum_{n=0}^{N-1} g[nM-k]e^{-j2\pi i\frac{n}{N}}\Delta t_w = \sum_{n_1=-k}^{M(N-1)-k} g[n_1]e^{-j2\pi f_i\times(n_1+k)\times\Delta t_w}\Delta t_w \approx \hat{g}(f_i)e^{-j2\pi i\frac{k}{NM}}$, where $\hat{g}(f)$ represents the discrete-time Fourier transform of the window function $g$. Substituting and pulling $\hat{g}(f_i)$ out of the summation yields,

$$I_N^{A_p}(f_i) \approx \left|\frac{\hat{g}(f_i)}{\sqrt{N}}\sum_k \epsilon_A[k]e^{-j2\pi i\frac{k}{NM}}\right|^2 = |\hat{g}(f_i)|^2 I_N^A(f_i)\ for\ large\ N \quad (5)$$

where $I_N^A(f_i)$ is the corresponding periodogram ordinate of true magnitude $A$ at the same frequency.

## III. Impact on Oscillation Detection

A generic oscillation detector is of the form $test\ stat > threshold$, where $test\ stat$ is the metric used to quantify oscillation severity. The performance of this detector depends on how well separated the distributions of its test statistic are under the null hypothesis (without oscillation) and the alternative hypothesis (with oscillation) (Fig. 1). This separability can be quantified using the area under the receiver operating characteristic (ROC) curve (AUC-ROC) [9], defined as,

$$AUC = \Pr(test\ stat|_{oscillation} > test\ stat|_{no\ oscillation}) \quad (6)$$

That is, the AUC is the probability that a randomly selected test statistic value during an active oscillation is higher than a

Chetan Mishra, Jaime Delaree Jr. and Kevin D. Jones are with Dominion Energy, VA, USA (chetan.mishra@dominionenergy.com)
Luigi Vanfretti is with Rensselaer Polytechnic Institute, NY, USA (vanfrl@rpi.edu)

randomly selected value under ambient, non-oscillatory conditions. The higher the AUC, the easier it is to construct a detector that detects oscillations while keeping the nuisance-alarm rate low.

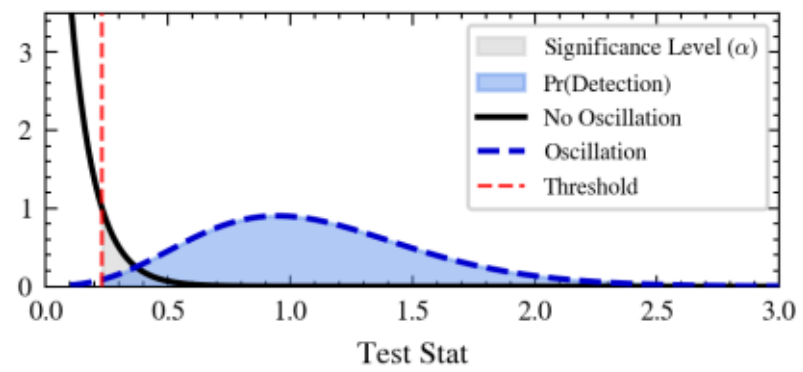

Fig. 1. Test-statistic distributions without/with oscillation, threshold, α, and detection probability.

In general, a detector using the ground-truth amplitude $A$ can be expected to outperform one based on the PMU estimate $A_p$, as the latter is inherently subject to DFT processing artifacts in addition to attenuation. To isolate the impact of attenuation, we evaluate whether detection performance degrades at more heavily attenuated frequencies. Specifically, we compare the AUC-ROC of two standard oscillation detectors identifying an oscillation of amplitude $A_o$ at a target frequency $f_o$ (DFT index $i_o$). The terms $AUC_{A_p}$ and $AUC_A$ denote the AUC of the detectors using the PMU-reported magnitude $A_p$ and the ground-truth magnitude $A$, respectively.

### A. Periodogram-Based Detector

Here, the test statistic [6] for detecting an oscillation at $f_o$ is the periodogram ordinate $I_N^{A_p}(f_o)$ itself. Substituting the periodogram relationship from (5) into (6),

$$AUC_{A_p} = \Pr\left(I_N^{A_p}(f_o)|_{A_o>0,f_o} > I_N^{A_p}(f_o)|_{A_o=0,f_o}\right) \quad (7)$$

$$= \Pr\left(|\hat{g}(f_o)|^2 I_N^A(f_o)|_{A_o>0,f_o} > |\hat{g}(f_o)|^2 I_N^A(f_o)|_{A_o=0,f_o}\right) \text{ for } |\hat{g}(f_o)| \neq 0$$

$$= \Pr\left(I_N^A(f_o)|_{A_o>0,f_o} > I_N^A(f_o)|_{A_o=0,f_o}\right) = AUC_A$$

Therefore, the performance of the periodogram-based detector is theoretically unaffected by DFT attenuation, i.e., it performs as well as with the unattenuated magnitude, unless $|\hat{g}(f_o)|^2 = 0$, in which case, $AUC$ collapses to 0.5 (pure chance).

### B. RMS Energy Detector

Here, the test statistic for detecting oscillations $\forall f_o \in B$ is the root-mean-square (RMS) energy [5] of $A_p$, denoted by $E_{RMS}^{A_p}(B)$. It can be expressed in terms of periodogram ordinates as [7],

$$E_{RMS}^{A_p}(B) = \sqrt{\frac{2}{N}\sum_{f_i \in B} I_N^{A_p}(f_i)} > 0 \quad (8)$$

Substituting (8) into (6) and applying (5) yields,

$$AUC_{A_p}(B) = \Pr\left(E_{RMS}^{A_p}(B)|_{A_o>0,f_o} > E_{RMS}^{A_p}(B)|_{A_o=0,f_o}\right) \quad (9)$$

$$= \Pr\left(\sum_{\substack{f_i \in B \\ f_i \neq f_o}} \beta_i^2 z_i \leq \kappa(A_o)\right)$$

$$AUC_A(B) = \Pr\left(\sum_{\substack{f_i \in B \\ f_i \neq f_o}} z_i \leq \kappa(A_o)\right)$$

where $z_i = \left(I_N^A(f_i)|_{A_o=0,f_o} - I_N^A(f_i)|_{A_o>0,f_o}\right)$, $\beta_i^2 = \frac{|\hat{g}(f_i)|^2}{|\hat{g}(f_o)|^2}$, the relative attenuation of $f_o$ with respect to $f_i$ and $\kappa(A_o) = \left(I_N^A(f_o)|_{A_o>0,f_o} - I_N^A(f_o)|_{A_o=0,f_o}\right)$. Thus, detection performance depends on the *attenuation* of the oscillation frequency relative to the other frequencies ($\beta_i \forall i$) in the same band, not on the absolute attenuation. Therefore, for a detection band over which the attenuation function $|\hat{g}(f)|$ is approximately constant (typically a narrow band), $\beta_i \approx 1 \forall i$ yields,

$$\textbf{\textit{Narrow}}: AUC_{A_p}(B) \approx \Pr\left(\sum_{\substack{f_i \in B \\ f_i \neq f_o}} 1 \times z_i \leq \kappa(A_o)\right) = AUC_A(B) \quad (10)$$

Thus, for narrow bands, detection performance using the PMU magnitude is approximately the same as that using the unattenuated ground truth. The periodogram detector is the special case in which the band contains a single DFT frequency. For the broad detection bands commonly used to capture higher-frequency grid dynamics, attenuation can vary significantly over the frequencies in $B$. To analyze this, we leverage the asymptotic statistical properties of periodograms for Gaussian processes $\epsilon_A[n]$ [6], [8] for a large sample size $N$,

$$I_N^A(f_i) \sim \frac{S_A(f_i)}{2}\chi_2^2(\lambda_i) \text{ where } \lambda_i = \begin{cases} \frac{A_o^2 N}{2S_A(f_o)}, & i = i_o \\ 0, & otherwise \end{cases} \quad (11)$$

$$I_N^A(f_1) \perp\!\!\!\perp I_N^A(f_2) \perp\!\!\!\perp \cdots$$

Here $\chi_2^2(\lambda_i)$ represents a noncentral chi-square distribution [10] with two degrees of freedom and $S_A(f)$ is the power spectral density of the ambient noise (in the absence of oscillation). Applying the central limit theorem [10] to approximate (9),

$$\textbf{\textit{Broad}}: AUC_{A_p}(B) \approx \Phi\left(\frac{\kappa(A_o)}{\sqrt{2\sum_{\substack{f_i \in B \\ f_i \neq f_o}} \beta_i^4 S_A^2(f_i)}}\right), \quad (12)$$

$$AUC_A(B) \approx \Phi\left(\frac{\kappa(A_o)}{\sqrt{2\sum_{\substack{f_i \in B \\ f_i \neq f_o}} S_A^2(f_i)}}\right)$$

Here, $\Phi$ is the cumulative distribution function of the standard normal distribution. Because $\kappa(A_o) > 0$ with high probability for large $N$, increasing the relative attenuation at the target frequency $f_o$ (increasing $\beta_i \forall i$) increases the denominator of $AUC_{A_p}$ in (12), thereby degrading the detector performance. However, neither the attenuated nor the unattenuated case outperforms the other in general. For instance, if $\beta_i > 1 \forall i \neq i_o$, the attenuated case (using PMU magnitude) can potentially outperform the baseline ($AUC_{A_p}(B) > AUC_A(B)$). This occurs because the non-oscillatory noise frequencies are attenuated more heavily than the actual oscillation frequency, thereby enhancing the signal-to-noise ratio.

## IV. Results

We construct a synthetic 60 Hz waveform with a mean amplitude $A(t)$ of 1 per unit (pu), sampled at 960 Hz. Ambient dynamics are modeled as band-limited ($< 60\,Hz$) Gaussian white noise, i.e., $S_A(f) = constant$. This isolates the effect of attenuation, which can then be assessed by comparing detection performance at different frequencies $f_o$. DFT-based synchrophasor estimates are obtained with a six-cycle boxcar window (within the nominal range of M-class PMUs [11]). Its frequency-domain attenuation function $|\hat{g}(f)|$ is shown in Fig. 2, along with the resulting magnitude $A_p$ and true amplitude $A$ spectra. Attenuation becomes noticeable above approximately 4 Hz. This window has spectral nulls at integer multiples of the inverse of its length, i.e., at multiples of 10 $Hz$. If attenuation degraded detection, performance would deteriorate progressively as the oscillation frequency increases from 0 to 10 Hz.

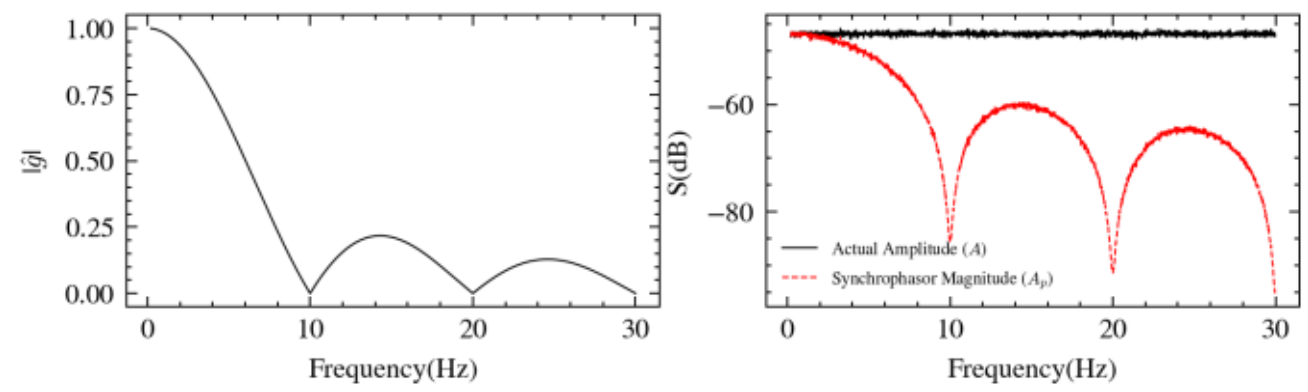

Fig. 2. Six-cycle boxcar window frequency response (left) and resulting attenuation of the magnitude spectrum (right).

We numerically validate the analytical results of Section III using 200 independent realizations of waveform data with and without oscillations at different $(A_o, f_o)$. The oscillation detection window duration $T$ is set to 60 s [6]. The two detectors (periodogram and RMS energy) are evaluated using two empirically estimated metrics: the probability of detection at a threshold that fixes the nuisance-alarm rate (significance level $\alpha$ [10]), estimated from non-oscillation samples (see Fig. 1), and the AUC, a threshold-independent measure of detectability.

For the periodogram-based detector, Fig. 3 shows the estimated probability of detection for $\alpha = 0.1$ [6] together with the estimated AUC. The detector's performance does not degrade at higher frequencies even though these are more strongly attenuated, as shown in Fig. 2, confirming the result of Section III-A. The exception is at fully attenuated frequencies ($|\hat{g}(f_o)| = 0$), e.g., 10 Hz here, where the detector is effectively blind (detection probability $\approx \alpha = 0.1$).

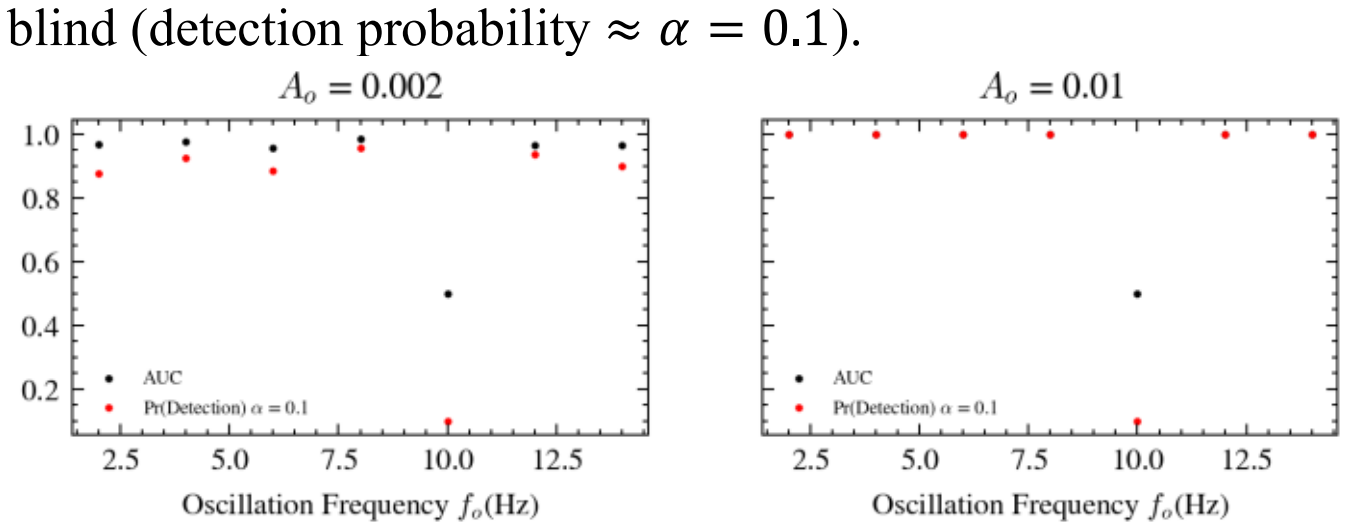

Fig. 3. Periodogram detector AUC and detection probability (α = 0.1) vs. $f_o$ for $A_o$= 0.002 (left) and 0.01 (right).

To evaluate the RMS energy detector, we use a wide detection band of $B = (1{,}15)$ Hz, which covers the 14.7 Hz data center oscillation reported in [1] and matches the upper edge of the 3–15 Hz high-frequency band used in PMU-based oscillation monitoring [12]; it also spans the local electromechanical (1–5 Hz) and high-frequency (above 5 Hz) bands monitored in practice [5]. As expected, the periodogram-based detector, a special case of the RMS energy detector, outperforms it, as reflected in Fig. 4's lower detection probabilities relative to Fig. 3 for the small amplitude $A_o = 0.002$. For the larger amplitude of 0.01, the performance roughly follows the trend of $|\hat{g}(f)|$ (Fig. 2), i.e., oscillations of equal amplitude are harder to detect at more attenuated frequencies, which confirms our derivation in Section III-B. For example, the probability of detecting an 8 Hz oscillation is below 20%, compared with almost 70% at 6 Hz. This degradation is relative to the other frequencies in the band and can largely be mitigated by using narrower detection bands, as predicted by (10). This is confirmed by the periodogram detector (the narrowest-band RMS energy detector), which has a detection probability of ≈ 100 % (with an AUC of ≈1) for the same 8 Hz oscillation.

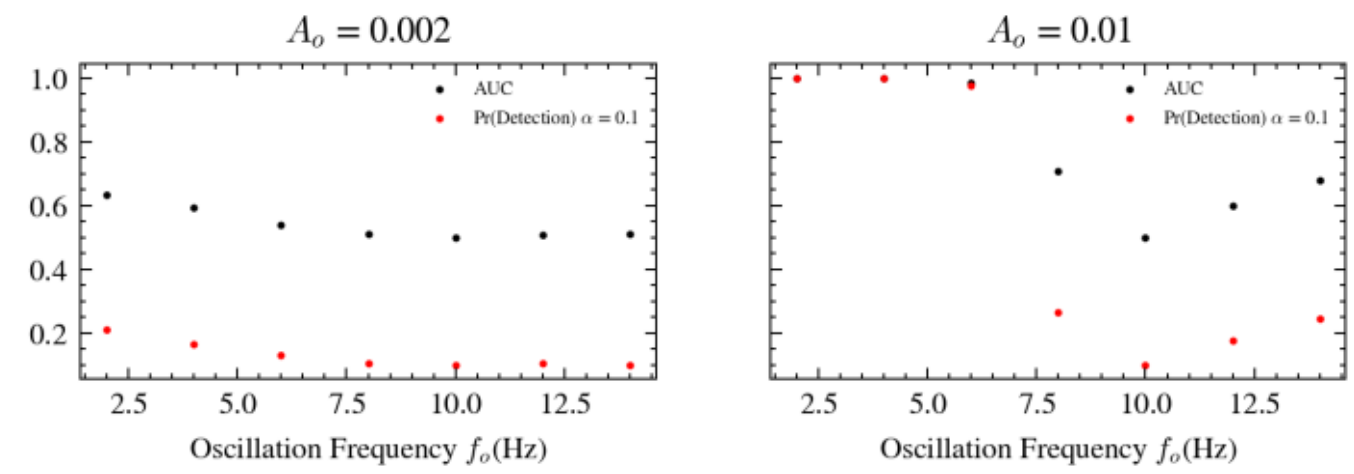

Fig. 4. RMS energy detector performance for $B = (1{,}15)\ Hz$.

## V. Conclusion

We demonstrate that the estimation window used in DFT-based synchrophasors scales the ground-truth periodogram by the squared magnitude of the window's frequency response. Furthermore, the performance of an RMS energy detector depends on the attenuation of the target oscillation frequency relative to other frequencies within the band, an effect that is largely mitigated by using narrow detection bands. For broad detection bands, we show that attenuation merely redistributes detectability across frequencies without a net loss, imposing no overall penalty on detection performance. Supported by Monte Carlo simulations, our results establish that standard synchrophasors can reliably detect high-frequency oscillations despite amplitude underestimation, provided that narrow-band RMS or, preferably, periodogram-based detectors are employed.

## References

[1] C. Mishra et al., "Understanding the inception of 14.7 Hz oscillations emerging from a data center," *Sustain. Energy Grids Netw.*, no. 101735, 2025.
[2] C. Wang et al., "Identifying oscillations injected by inverter-based solar energy sources," in *Proc. IEEE PES GM*, 2022.
[3] B. Ou et al., "Applicability and limitation analysis of PMU data and phasor concept for low- and high-freq. oscillations," arXiv:2601.15529, 2026.
[4] K. Chatterjee et al., "Measurement adequacy for monitoring data center oscillations," PNNL, Richland, WA, USA, Tech. Rep. PNNL-39191, 2026.
[5] M. Donnelly et al., "RMS-energy filter design for real-time oscillation detection," in *Proc. IEEE PES GM*, 2015.
[6] J. Follum and J. W. Pierre, "Detection of periodic forced oscillations in power systems," *IEEE Trans. Pwr. Syst.*, vol. 31, no. 3, pp. 2423–2433, 2016.
[7] J. Follum et al., "A statistics-based threshold for the RMS-energy oscillation detector," *Int.J.Electr.Pwr. Energy Syst.*, v. 128, no. 106685, 2021.
[8] M. B. Priestley, *Spectral Analysis and Time Series*. London, U.K., 1981.
[9] T. Hastie et al., *The Elements of Statistical Learning*, 2nd ed, Springer, 09.
[10] G. Casella and R. L. Berger, *Statistical Inference*. Boca Raton, FL, USA: CRC Press, 2024.
[11] *Measuring Relays and Protection Equipment—Part 118-1: Synchrophasor for Power Systems—Measurements*, IEEE/IEC Std 60255-118-1:2018, 2018.
[12] L. Vanfretti et al., "A phasor measurement unit based fast real-time oscillation detection application for monitoring wind-farm-to-grid sub-synchronous dynamics," *Electr. Pwr. Compon. Syst.*, vol. 44, no. 2, 2016.